\documentclass{article}
\usepackage{spconf,amsmath,graphicx}

\usepackage{enumitem}
\setlist{nosep, leftmargin=14pt}
\let\svthefootnote\thefootnote
\newcommand\freefootnote[1]{%
  \let\thefootnote\relax%
  \footnotetext{#1}%
  \let\thefootnote\svthefootnote%
}
\usepackage[utf8]{inputenc}
\usepackage[english]{babel}
\usepackage{cite}
\usepackage{graphicx}
\usepackage{hyperref}

\usepackage{amsfonts} 
\usepackage{float}
\usepackage{xcolor}
\usepackage{caption}
\usepackage{subcaption}
\newcommand{\cmmnt}[1]{\ignorespaces}
\usepackage{eqnarray,amsmath}
\usepackage{amsmath,bm}
\usepackage{amssymb}
\usepackage{graphicx}
\usepackage[linesnumbered,ruled,vlined]{algorithm2e}

\title{ULTRASOUND ELASTICITY IMAGING USING PHYSICS-BASED MODELS AND LEARNING-BASED PLUG-AND-PLAY PRIORS}
\name{%
    Narges Mohammadi$^{\star}$
    \qquad Marvin M. Doyley$^{\star}$%
    \qquad Mujdat Cetin$^{\star}{}^{\dagger}$}
\address{$^{\star}$ Department of Electrical and Computer Engineering, University of Rochester, Rochester, NY, USA \\%
   $^{\dagger}$ Goergen Institute for Data Science, 
University of Rochester, Rochester, NY, USA
}

\begin{document}

\maketitle

\begin{abstract}
Existing physical model-based imaging methods for ultrasound elasticity reconstruction utilize fixed variational regularizers that may not be appropriate for the application of interest or may not capture complex spatial prior information about the underlying tissues. On the other hand, end-to-end learning-based methods count solely on the training data, not taking advantage of the governing physical laws of the imaging system. Integrating learning-based priors with physical forward models for ultrasound elasticity imaging, we present a joint reconstruction framework which guarantees that learning driven reconstructions are consistent with the underlying physics. For solving the elasticity inverse problem as a regularized optimization problem, we propose a plug-and-play (PnP) reconstruction approach in which each iteration of the elasticity image estimation process involves separate updates incorporating data fidelity and learning-based regularization. In this methodology, the data fidelity term is developed using a statistical linear algebraic model of quasi-static equilibrium equation revealing the relationship of the observed displacement fields \cmmnt{measured deformation data} to the unobserved elastic modulus. The regularizer comprises a convolutional neural network (CNN) based denoiser that captures the learned prior structure of the underlying tissues. Preliminary simulation results demonstrate the robustness and effectiveness of the proposed approach with limited training datasets and noisy displacement measurements.
\end{abstract}
\section{Introduction}
\freefootnote{This work has been partially supported by the National Science Foundation (NSF) under Grant CCF-1934962.}
Ultrasound elasticity imaging has significant potential in tissue stiffness quantification \cmmnt{and characteristic visualization} enabling reliable diagnostic decisions\cmmnt{clinical diagnosis}. Elasticity reconstruction problem which is cast as solving an ill-posed inverse problem can be formulated as a regularized optimization task that involves a forward model describing physics of data acquisition and regularization constraints describing the prior information about the latent image. For addressing medical imaging concerns, many approaches have been developed for both fast elasticity imaging as well as robust and accurate image reconstruction using limited noisy deformation measurements.
Existing model-based approaches in elasticity imaging \cite{Marvin}, \cite{asilomar} typically assume fixed regularization terms for various tissue types in elastography tasks while advanced priors might be required to mitigate the corresponding corrupted incomplete measurements and also to capture complex spatial information about the underlying tissues.  \\
On the other hand, end-to-end deep learning methods \cite{E2E,E2E2,E2E3,shear} require large datasets which conflict with fast and time-efficient image reconstruction essentials by real-time medical applications. Moreover, as the forward measurement model is not explicitly used in such methods, the estimated solution may not be consistent with the physics governing the imaging problem. \cmmnt{One other challenge of using end-to-end DNN is that retraining the network for different tasks is expensive which means whenever the forward model, noise distribution or noise level change, a new network is required to be trained which prevents obtaining to a high level of generalizability \cmmnt{and require the retraining the network for each specific problem}.}\\
These limitations can be suppressed by integrating the \cmmnt{(forward model and learned priors)} physical models and the learning-based priors as the complementary information sources to enable fast and accurate elasticity imaging \cite{willet,PGD,selfsuper}. There are two types of approaches for combining model-based and learning-based methods.
The first group of methods is based on unrolling concepts of the classical optimization iterations where each iteration is a layer of a neural network involving the forward model. These types of approaches including physics-informed learning methods (\cite{maziar},\cite{PI-DL}, PINN \cite{PINN}, PI-GAN \cite{GAN} and MoDL \cite{MoDL}) provide improved accuracy while they are time-consuming as they require network retraining at each iteration which makes them impractical for some medical applications. \\
The second group of approaches combine learned regularizers and physics knowledge within the framework of model-based image reconstruction. Most of these methods including plug-and-play prior (PnP) \cite{PnP_CT} and regularization by denoising (RED \cite{RED}, N2N \cite{N2N}) \cmmnt{and Regularization by artifact removal (RARE \cite{RARE})} solve the imaging optimization problem by separating the forward model from the learned prior. Once the regularization network is trained, it can be integrated into the iterations of reconstruction procedure such as alternating direction method of multipliers (ADMM) and other proximal splitting methods \cite{proximal}, \cite{denoiser_prior}. One major benefit of these approaches is that they promote generalization by plugging any forward model independent of the regularization term. \cmmnt{update and this feature is beneficial when sufficient training pairs are available while training them for the distinct forward operator is expensive.
if the forward model is available at the test time, the learned network using training pairs can be exploited for any forward model}\\
In this paper, we propose a statistical learning approach based on PnP methodology for ultrasound elasticity image reconstruction in which the elasticity image is estimated using data fidelity and learned regularization term updates. We model the data fidelity term using the equilibrium equation of elasticity as a linear representation for elasticity modulus and learn a CNN-based denoiser to be employed as the proximal operator of the regularizer in an iterative reconstruction approach. Taking advantage of feasible computing of the gradient of data fidelity term and learning the proximal operator of the regularizer encourages employing the proximal gradient methods which lead to robust and interpretable elasticity reconstruction. Our simulation results verify the effectiveness of the proposed PnP methodology with limited training datasets and noisy displacement fields.\\
The remainder of this manuscript is organized as follows. We analyze the inverse problem formulation for ultrasound elasticity imaging in Section \ref{sec2}. The PnP approach and the proposed paradigm are elaborated in Section \ref{sec3}. The simulation results of elasticity image reconstruction are presented in Section \ref{sec4}, and finally, concluding remarks are provided in Section \ref{sec5}. 
\cmmnt{
(PnP is a nonconvex framework that integrate modern learning-based priors into proximal operator to improve the optimization performance.) (Training the proximal operator is performed distinctly from any forward model.) By benefiting explicit regularizer, we can substitute the denoiser with a (more general image restoration function)
}
\section{Inverse Problem Formulation}
\label{sec2}
The forward model of ultrasound elasticity imaging for incompressible tissue with plane strain assumption is governed by quasi-static equilibrium equation\cmmnt{described over the discretized cross-section of the tissue}. The statistical formulation of this equilibrium condition known as global stiffness equation which relates the measured deformation fields $\mathbf{u}$ and force vector $\mathbf{f}$ to the unknown elasticity modulus $\mathbf{E}$ can be presented as:
\begin{equation}
\label{eq:1}
\mathbf{f}=\mathbf{K}(\mathbf{E})\mathbf{u}+\mathbf{w}\qquad \mathbf{w}\sim \mathcal{N}(0,\,\bm{\Sigma_{w}})
\end{equation}
Letting $\mathbf{N}$ denote the number of mesh nodes, $\mathbf{f}\in \mathbb{R}^{2N\times 1}$ represents the global nodal force measurements including boundary conditions in lateral and axial directions, $\mathbf{u}\in \mathbb{R}^{2N\times 1}$ denotes the noiseless global nodal deformation measurements, and $\mathbf{w}\in \mathbb{R}^{2N\times 1}$ represents the nodal Gaussian noise fields. $\mathbf{K}(\mathbf{E})\in  \mathbb{R}^{2N\times 2N}$ relates force and deformation fields as a function of tissue elasticity distribution $\mathbf{E}\in \mathbb{R}^{N\times 1}$. The inverse problem of estimating the elasticity modulus $\mathbf{E}$ can be formulated as a regularized optimization problem. To this end, it is required to reformulate the forward model (\ref{eq:1}) by extracting the latent elasticity modulus from the global stiffness matrix \cmmnt{ from the physical measured system} as the unknown vector. In this respect, we introduce the matrix $\mathbf{D}(\mathbf{u})\in \mathbb{R}^{2N\times N}$ \cite{FEM} which is related to $\mathbf{K}(\mathbf{E})$ by a 3D tensor $\bm{\Psi}\in \mathbb{R}^{N\times 2N\times2N}$ developed from the equilibrium equation and by applying Poisson's ratio $\nu$ and Neumann boundary conditions:
\begin{equation}
\label{eq:2}
\mathbf{D}(\mathbf{u})\mathbf{E}=\mathbf{K}(\mathbf{E})\mathbf{u}\\
\end{equation}
\begin{equation}
\label{eq:3}
\mathbf{D}(\mathbf{u})=(\bm{\Psi}\mathbf{u})^{T} \qquad \mathbf{K}(\mathbf{E})=\bm{\Psi}^{T}\mathbf{E}
\end{equation}
The displacement fields are acquired by cross-correlation of multiples B-mode ultrasound images which introduce the noisy displacement observations as $\mathbf{u^{m}}=\mathbf{u}+\mathbf{n}$ where $\mathbf{n}\sim \mathcal{N}(0,\,\bm{\Sigma_{n}})$. Integrating this observation process into the statistical forward model (\ref{eq:1}) results in:
\begin{eqnarray}
\label{eq:5}
\mathbf{f}&=&\mathbf{K}(\mathbf{E})\mathbf{u}+\mathbf{w}=\mathbf{K}(\mathbf{E})(\mathbf{u^{m}}-\mathbf{n})+\mathbf{w}\nonumber\\
&=&\mathbf{K}(\mathbf{E})\mathbf{u^{m}}-\mathbf{K}(\mathbf{E})\mathbf{n}+\mathbf{w}
\end{eqnarray}
Letting ${\mathbf{\Tilde{w}}}=-\mathbf{K}(\mathbf{E})\mathbf{n}+\mathbf{w}$ and employing (\ref{eq:2}) using noisy displacements, $\mathbf{D}(\mathbf{u^{m}})\mathbf{E}=\mathbf{K}(\mathbf{E})\mathbf{u^{m}}$, the unified statistical forward model could be described as:
\begin{equation}
\label{eq:6}
\mathbf{f}=\mathbf{D}(\mathbf{u^{m}})\mathbf{E}+\mathbf{\Tilde{w}}\qquad \mathbf{\Tilde{w}}\sim \mathcal{N}(0,\,\bm{\Gamma})
\end{equation}
where $\bm{\Gamma}$ is computed by:

\begin{equation}
\label{eq:7}
\bm{\Gamma}=\bm{\Sigma_{w}}+\mathbf{K}(\mathbf{E})\bm{\Sigma_{n}}\mathbf{K}(\mathbf{E})^{T}
\end{equation}
This underlying forward model enables us to formulate the elasticity inverse problem as a constrained optimization problem given by:

\begin{equation}
\label{eq:8}
\begin{array}{l}
\mathbf{\hat{E}}=\mathrm{argmin} _{\mathbf{E}}\quad\frac{1}{2}\left \|  \mathbf{f}-\mathbf{D}(\mathbf{u^{m}})\mathbf{E} \right \|_{{\bm{\Gamma}}^{-1}}^{2}+\lambda R(\mathbf{E})\\
\quad\quad\quad
s.t.\quad \mathbf{E}>0
\end{array}
\end{equation}
where $\left \| \mathbf{A} \right \|_{\mathbf{B}}^{2}:=(\mathbf{A}^{T}\mathbf{B}\mathbf{A})$ and $R$ represents the regularizer. The optimization problem in (\ref{eq:8}) introduces a  new representation for the elasticity inverse problem, which can be solved by the use of a fixed-point procedure, fixing $\bm{\Gamma}$ during the update of $\mathbf{E}$, and then updating $\bm{\Gamma}$ by plugging the new estimate of $\mathbf{E}$ in (\ref{eq:7}). For updating the elastic modulus $\mathbf{E}$, we use proximal gradient algorithm \cite{proximal} as follows:
\begin{equation}
\label{eq:9}
\mathbf{E}_{n+1}=\textrm{prox}_{\mathbf{E}_{n}>0}(\textrm{prox}_{R\lambda\gamma_{n}}(\mathbf{E}_{n}-\gamma _{n}\nabla g(\mathbf{E}_{n})))
\end{equation}
Furthermore, one can compute $\nabla g(\mathbf{E})$ using:
\begin{equation}
\label{eq:10}
g(\mathbf{E})=\frac{1}{2}(\mathbf{f}-\mathbf{D}(\mathbf{u^{m}})\mathbf{E}  )^{T}\bm{\Gamma}^{-1}(\mathbf{f}-\mathbf{D}(\mathbf{u^{m}})\mathbf{E}  )
\end{equation}
\begin{equation}
\label{eq:11}
\nabla g(\mathbf{E})=-(\mathbf{D}(\mathbf{u^{m}}))^{T}\bm{\Gamma}^{-1}(\mathbf{f}-\mathbf{D}(\mathbf{u^{m}})\mathbf{E} )
\end{equation}
According to the elasticity update formulation (\ref{eq:9}), the proximal gradient approach decouples the data fidelity term update and the proximal operator of the regularizer which facilitates the PnP methodology for applying a learning-based \cmmnt{data-driven} prior.  

\section{LEARNING-BASED Plug-and-play prior methodology}
\label{sec3}
\begin{figure}[t]
  \centering
  \centerline{\includegraphics[width=8cm]{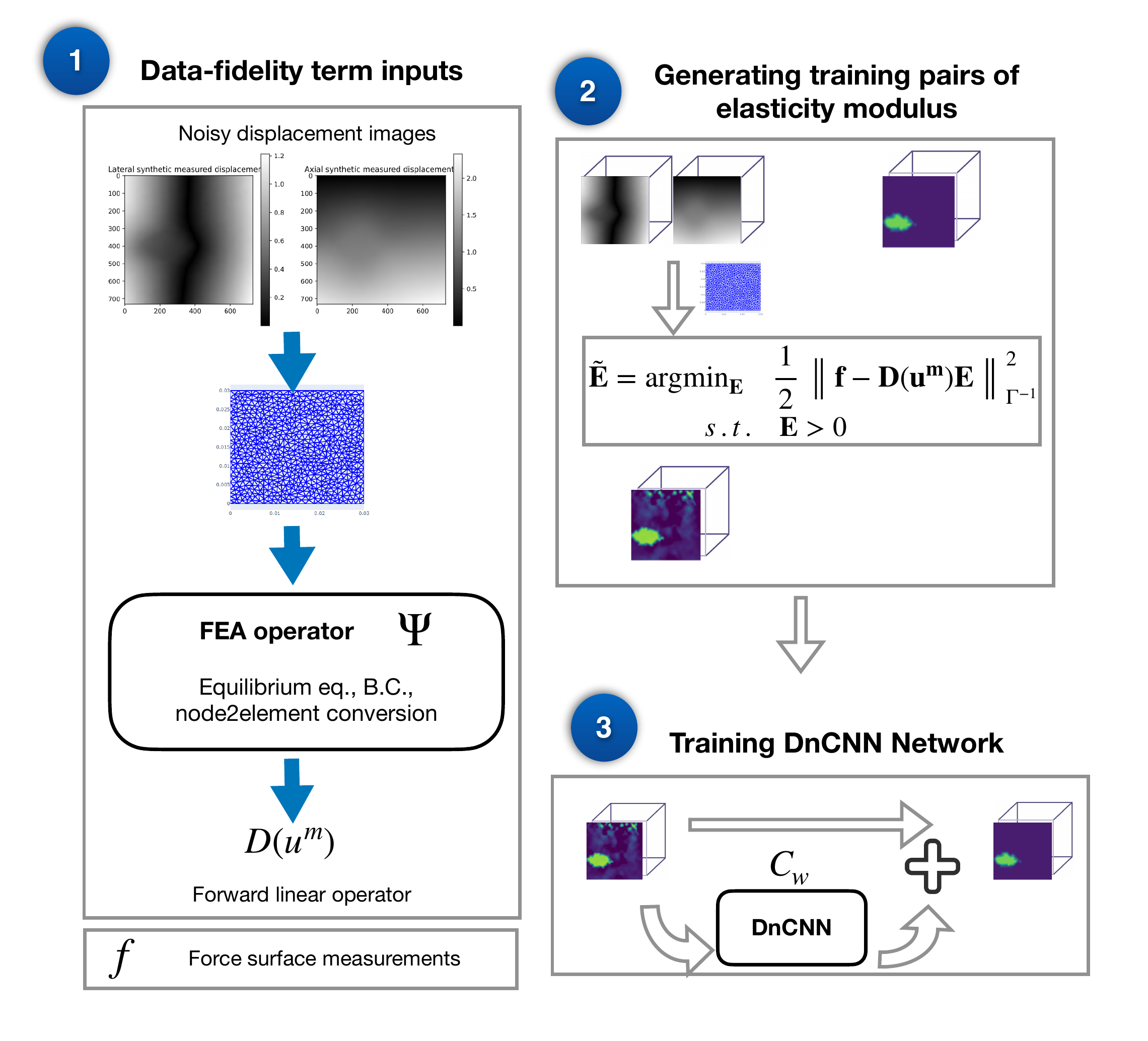}}
\caption{Complete training procedure. In steps 1 and 2, training elasticity pairs are generated. In step 3, the denoiser network parameters are learned by feeding the noisy elasticity images and the clean ones into the network.}
\label{fig:1}
\end{figure}

\begin{figure}[h!]
  \centering
  \centerline{\includegraphics[width=9cm]{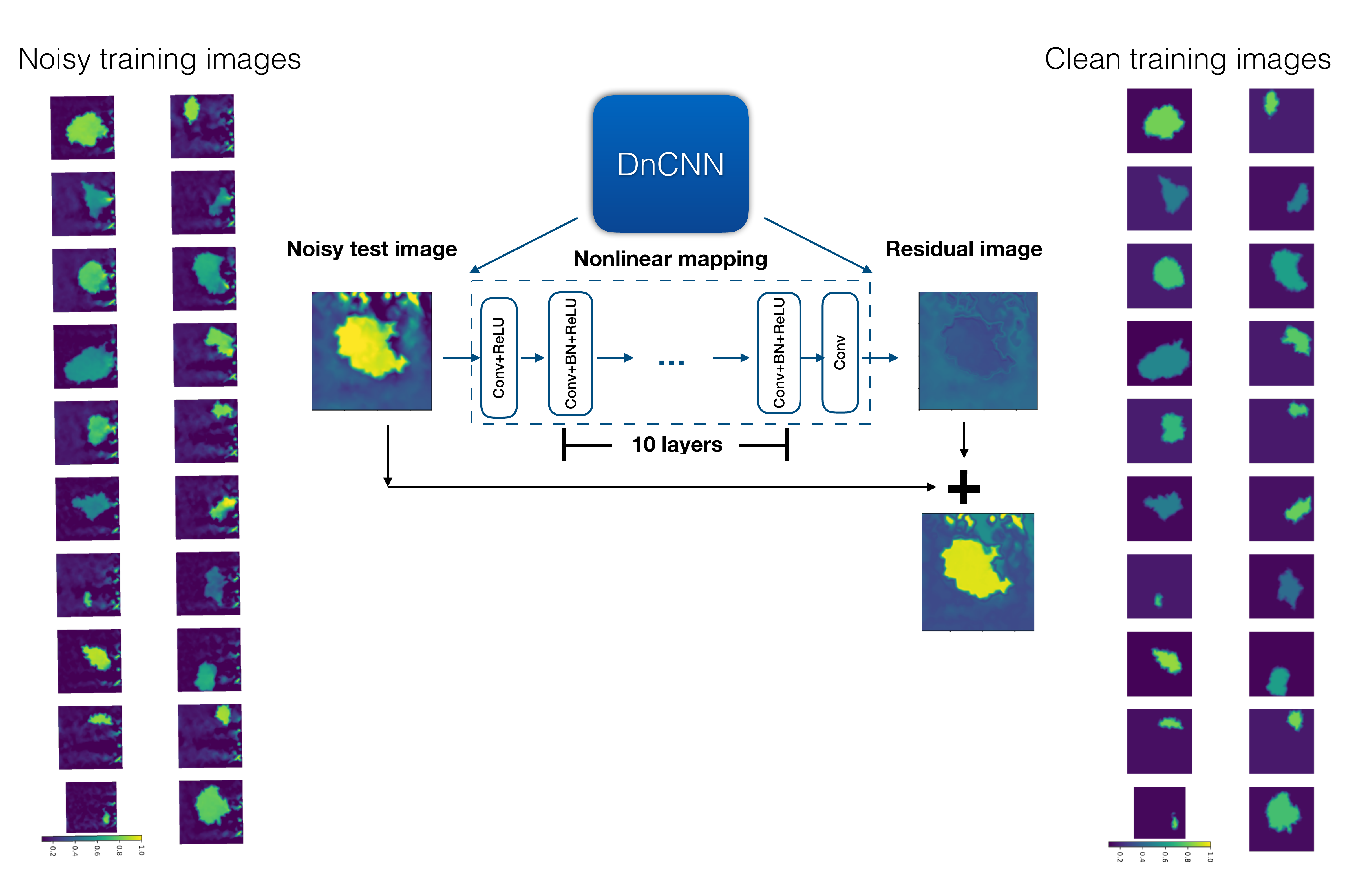}}
\caption{The training procedure of the denoiser network using noisy and clean elasticity images.}
\label{fig:3}
\end{figure}

\begin{figure}[t]
  \centering
  \centerline{\includegraphics[width=8cm]{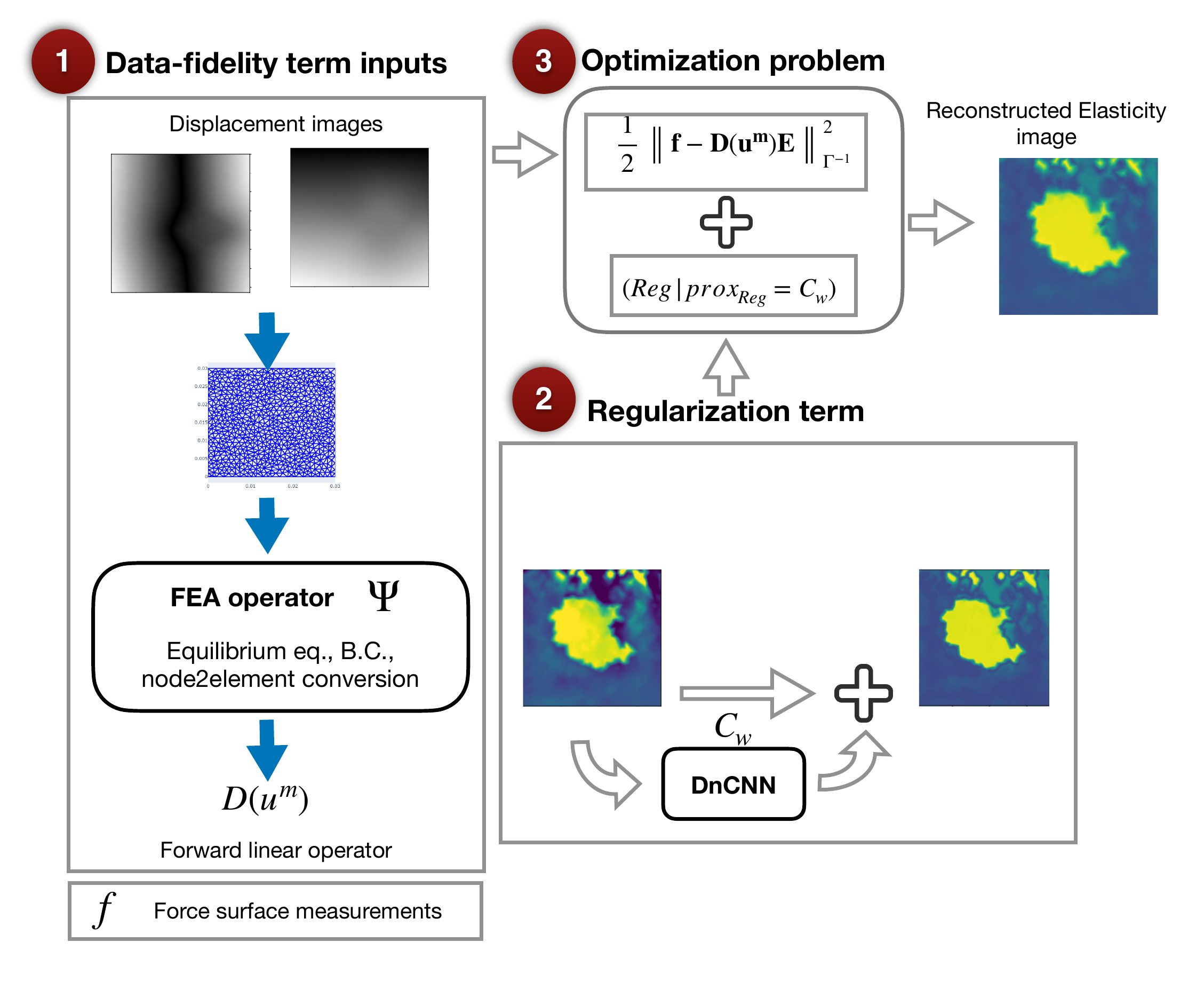}}
\caption{Learning-based elasticity reconstruction of a test image using the proposed learning-based PnP method.}
\label{fig:2}
\end{figure}

\cmmnt{In this section, we introduce a new elasticity image reconstruction framework that benefits from alternating minimization algorithms using projection to (convex) manifolds which leads to guaranteed reconstruction, reduced computation time, and reduced training dataset size.}
For estimating the elasticity modulus $\mathbf{E}$, the prior information of the latent images should be applied to (\ref{eq:9}) as the regularizer. Here, we examine the use of learning-based advanced priors with the goal of capturing complex spatial structures of underlying tissues. PnP approaches allow us to extract data-driven prior information in elasticity images by supervised learning \cmmnt{/unsupervisedly/generative methods} and apply this learned prior as the proximal operator of the regularizer in (\ref{eq:9}) independent of the data fidelity term. \\
The overall training procedure is illustrated in Fig.\ref{fig:1}. First, noisy displacement images are discretized over the mesh elements and the linear forward operator $\mathbf{D(u^{m})}$ is constructed for applying to the data-fidelity term. Next, the noisy training sets $\tilde{\mathbf{E}}$ are generated by benefiting from \textit{maximum likelihood} (ML) estimates of elasticity modulus from poor noisy displacement measurements $\mathbf{u^{m}}$ without any priors. 
Finally, to achieve a data-driven regularizer, we train a denoiser network based on the DnCNN \cite{DnCNN} model, $C_{w}(\tilde{\mathbf{E}})$ parameterized with weights $w$, by learning the residuals between noisy and clean elasticity training pairs. \cmmnt{noisy elasticity images $\tilde{\mathbf{E}}$ and clean elasticity images.}
\cmmnt{which could be described as first mapping the noisy displacement measurements $\mathbf{u^{m}}$ to the ML estimation of the elasticity images \cmmnt{using the inverse of the forward operator} which leads to poor noisy elasticity reconstructions $\tilde{\mathbf{E}}$ and then training the denoiser network using these noisy elasticity images and the corresponding clean reference images $\mathbf{E}$. }It is worth noting that the noise present in the input images is beyond Gaussian, modeled as signal-dependent colored noise with the covariance matrix $\bm{\Gamma}$ described in (\ref{eq:7}). The details of the denoiser network training (which is step 3 in Fig.\ref{fig:1}) is elaborated in Fig.\ref{fig:3}. The network output is the residual of clean and noisy elasticity images \cmmnt{to prevent vanishing gradients during the training by} using the following mean squared error (MSE) loss function \cite{PnP_CT}: 
\begin{equation}
\label{eq:13}
\emph{l}(w)=\frac{1}{2N}\sum^{N}\left \| C_{w}(\tilde{\mathbf{E}})-(\tilde{\mathbf{E}}-\mathbf{E}) \right \|_{F}^{2}
\end{equation}
The PnP reconstruction framework is depicted in Fig.\ref{fig:2} which illustrates that the learned regularizer $C_{w}(\tilde{\mathbf{E}})$ is plugged into (\ref{eq:9}) as the proximal operator of the learning-based regularizer \cite{tum}. Therefore, the iterative estimation procedure of the elasticity modulus can be described as:
\begin{equation}
\label{eq:12}
\mathbf{E}_{n+1}=\textrm{prox}_{\mathbf{E}_{n}>0}(C_{w}(\mathbf{E}_{n}-\gamma _{n}\nabla g(\mathbf{E}_{n})))
\end{equation}
It is worth mentioning that \cmmnt{integrating the learning-based prior to the forward model update in (\ref{eq:12}), guarantees that the projection of the learned prior lies in the manifold of the governing physical model. Moreover,} this approach does not require learning the physical system, as the statistical forward model is known; therefore, the network parameters are not wasted on learning the physical model, leading to a reduced size of dataset requirement. These benefits verify that the PnP method enables robust elasticity reconstructions in reduced computation time.

\section{EXPERIMENTAL RESULTS}
\label{sec4}
\begin{figure}[t]
\begin{minipage}[b]{.45\linewidth}
  \centering
  \centerline{\includegraphics[width=2.8cm]{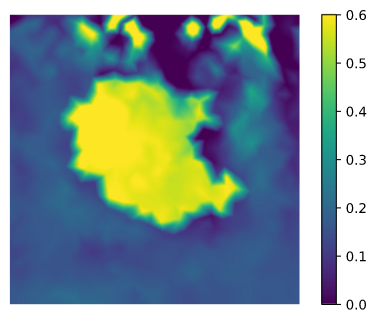}}
    \vspace{-0.6\baselineskip}
  \centerline{(a) \scriptsize{}}\medskip
\end{minipage}
\begin{minipage}[b]{0.45\linewidth}
  \centering
  \centerline{\includegraphics[width=2.8cm]{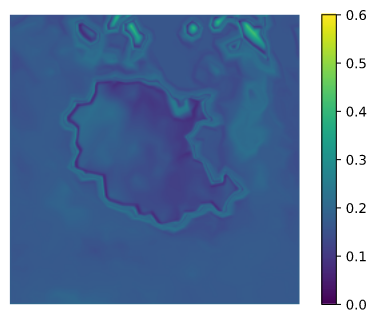}}
    \vspace{-0.6\baselineskip}
  \centerline{(b) \scriptsize{}}\medskip
\end{minipage}
\par\vspace{+0.25\baselineskip}
\begin{minipage}[b]{0.45\linewidth}
  \centering
  \centerline{\includegraphics[width=2.8cm]{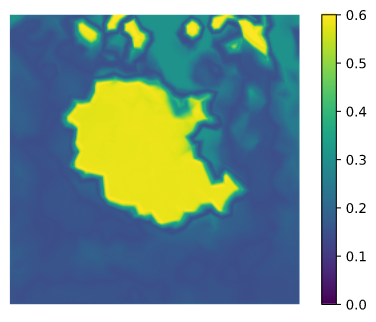}}
    \vspace{-0.6\baselineskip}
  \centerline{(c) \scriptsize{}}\medskip
\end{minipage}
\begin{minipage}[b]{0.45\linewidth}
  \centering
  \centerline{\includegraphics[width=2.8cm]{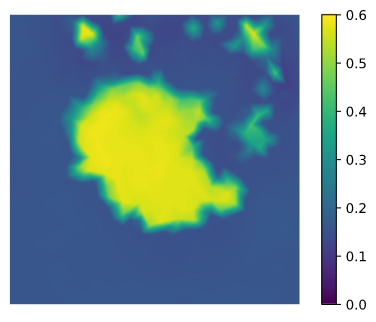}}
    \vspace{-0.6\baselineskip}
  \centerline{(d) \scriptsize{}}\medskip
\end{minipage}
\par\vspace{+0.25\baselineskip}
\begin{minipage}[b]{0.45\linewidth}
  \centering
  \centerline{\includegraphics[width=2.8cm]{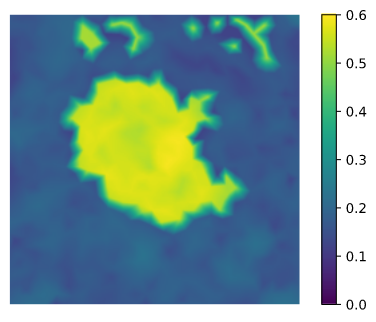}}
    \vspace{-0.6\baselineskip}
  \centerline{(e) \scriptsize{}}\medskip
\end{minipage}
\begin{minipage}[b]{0.45\linewidth}
  \centering
  \centerline{\includegraphics[width=2.8cm]{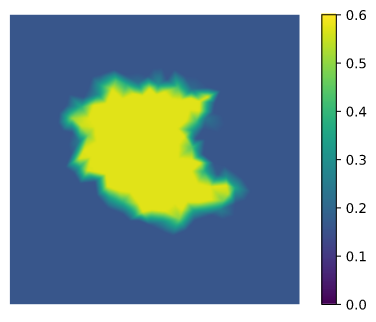}}
    \vspace{-0.6\baselineskip}
  \centerline{(f) \scriptsize{}}\medskip
\end{minipage}
\caption{(a) Estimated elasticity image without regularization term which is the input test image to the trained network. (b) residual image as the output of the trained network. (c) reconstructed elasticity image with learning-based post-processing. (d) reconstructed elasticity image with the proposed PnP method. (e) reconstructed elasticity image with total variation (TV) regularizer. (f) ground-truth elasticity image. The unit of the color bar is 100 KPa. }
\label{fig:4}
\end{figure}

\begin{figure}[h!]
  \centering
  \centerline{\includegraphics[width=5cm]{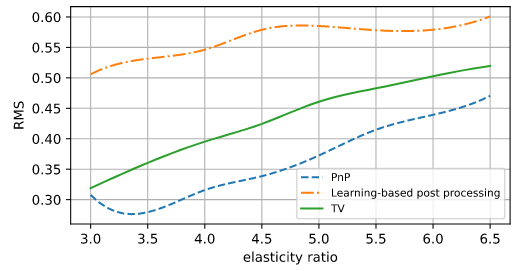}}
\caption{RMS performance metric achieved by the proposed PnP approach, the TV-regularized one, and the learned post-processing approach for a range of inclusion to background elasticity ratio. SNR=35dB. }
\label{fig:5}
\end{figure}

The main step in solving the elasticity imaging problem using the proposed framework is generating the training data pairs. \cmmnt{Commonly, the available measurements for ultrasound elasticity imaging problem includes noisy displacement fields $\mathbf{u^m}$ and force boundary conditions $\mathbf{f}$. It is required to map these measurement fields to the elasticity image domain for training the denoiser network as the proximal operator of the regularizer. In this regard, we use a training dataset consisting of 541 clean elasticity images $\mathbf{E}$ of breast tissues and the equivalent displacement images $\mathbf{u}$. }In this regard, 541 B-mode images of real breast lesion presented in \cite{dataset} are used to create the true elasticity images (synthetic $\mathbf{E}$ maps). To this aim, the elasticity of the lesion and background are chosen as random scales by constraining that normalized background elasticity is in range of 0.1-0.15 KPa and normalized lesion elasticity is in range of 0.3-0.8 KPa; therefore, the ratio of lesion elasticity to the background elasticity falls in range of 2-8  following the experimental records. Moreover, the displacement images $\mathbf{u^{m}}$ are computed for each $\mathbf{E}$ image by solving the forward model. For simulating more realistic measurements, multivariate Gaussian noise with $SNR=35dB$ is added to the displacement images, and noisy elasticity images $\mathbf{\tilde{E}}$ are reconstructed by ML estimation without any regularizer (only positivity constraint is applied). These noisy elasticity images $\mathbf{\tilde{E}}$ along with the clean elasticity images $\mathbf{E}$, both with the size of $730\times 730$, are fed into the aforementioned DnCNN depicted in Fig.\ref{fig:3} consisting of convolution layers with a kernel size of $3\times 3$ each of which is followed by a rectified linear unit (ReLU) for residual learning. For selecting network parameters, our simulation results indicate that by increasing the number of layers to 20, we achieve higher accuracy in inclusion geometry and shape estimation while we lose the accuracy in the quantitative value of elasticity modulus estimation; thus, we set the number of layers to 10 to have less reconstruction error. Other detailed settings of the network can be described as patch-size=50, batch-size=16, num-epochs=20, and learning-rate=1e-4.
To evaluate the performance of the proposed PnP approach, we compare it with a model-based approach with total variation (TV) regularizer and a learning-based denoiser applied as post-processing after standard image formation. We call this approach learned post-processing. For a particular test sample, the reconstructed images, network input test image, the residual image, and the ground truth image are shown in Fig.\ref{fig:4}. For quantitative performance evaluation, we compute  RMS (relative mean squared) error as a performance metric for the PnP approach, model-based TV-regularized approach, and learning-based post-processing one over the test dataset consisting of 100 images. The simulation results illustrated in Fig.\ref{fig:5} indicates that the PnP method achieves the smallest RMS error over all elasticity ratios compared to model-based TV approach and learned post-processing one. On the other hand, the computation time of the TV-regularized approach is nearly 6.1 times of the PnP method and the learned post-processing approach requires only 0.11 times the computation time of the PnP method. Considering the accuracy and speed of proposed PnP approach compared to two other methods demonstrate the effectiveness of the PnP for elasticity reconstruction.
\cmmnt{The RMS results indicate that although the learned denoiser provides high CNR for large rations of elasticity playing, the estimated images are not compatible with the true ones.}

\section{Conclusion}
\label{sec5}
This article presents a PnP methodology for solving the inverse problem in ultrasound elasticity imaging by combining statistical model-based representations and learning-based regularizers. We derive an integrated optimization function and benefit from proximal gradient methods as its solver which decouple the iterative optimization steps into the data fidelity and regularizer updates. DnCNN is used for exploiting the underlying prior information which is plugged into the optimization function as the regularizer proximal operator to ensure that the reconstructed images are projected to the statistical physical model. Our simulation results demonstrate the improved image reconstruction efficiency in terms of quality and computation time.
\bibliographystyle{IEEEbib}
\bibliography{Mybib}

\end{document}